\documentclass[prl,superscriptaddress,reprint,showpacs,amsmath,amssymb,nofootinbib,nobalancelastpage]{revtex4-1}
\usepackage{amsmath,amssymb,graphicx,verbatim}
\usepackage{bbold,color} 
\usepackage[colorlinks=true,urlcolor=blue,linkcolor=blue]{hyperref}

\font\tenbsl=cmbxsl10

\def\({\left(}
\def\){\right)}

\def\a2d{a^{\dagger 2}}
\def\b2d{b^{\dagger 2}}

\def\eq#1{Eq.~(\ref{eq:#1})}
\def\Eq#1{Equation~(\ref{eq:#1})}

\def\fig#1{Fig.\ref{fig:#1}}

\def\Fig#1{Figure~\ref{fig:#1}}

\def\tab#1{Table~\ref{tab:#1}}

\def\tr#1#2{\mathrm{Tr}_{#1}\left[#2\right]}

\newcommand\bra[1]{\left\langle\,#1\,\right|} 
\newcommand\ket[1]{\left|\,#1\,\right\rangle}

\newcommand\om[1]{\omega_{\scriptscriptstyle #1}}

\def\og{\leavevmode\raise.3ex\hbox{$\scriptscriptstyle\langle\!\langle$}}
\def\fg{\leavevmode\raise.3ex\hbox{$\scriptscriptstyle\,\rangle\!\rangle$}}

\def\refer#1#2{[{\tenbsl{#1}
\setbox100=\hbox{#2}\ifdim\wd100>10pt\kern .3em\box100$\,$\fi}]}

\def\lejour{le\ {\the\day}\
\ifcase\month\or janvier\or f\'evrier\or mars\or avril\or mai\or juin\or
juillet\or ao\^ut\or septembre\or octobre\or novembre\or d\'ecembre\fi\
{
\the\year}}

\def\boxit#1#2{\setbox1=\hbox{\kern#1{#2}\kern#1}%
\dimen1=\ht1 \advance\dimen1 by #1 \dimen2=\dp1 \advance\dimen2 by #1
\setbox1=\hbox{\vrule height\dimen1 depth\dimen2\box1\vrule}%
\setbox1=\vbox{\hrule\box1\hrule}%
\advance\dimen1 by .4pt \ht1=\dimen1
\advance\dimen2 by .4pt \dp1=\dimen2 \box1\relax}

\begin{document}
\title{State-independent quantum tomography of a single-photon state by photon-number-resolving measurements}
\email{Contribution of NIST, an agency of the U.S. government, not subject to copyright.}
\author{Rajveer Nehra}
\email{rn2hs@virginia.edu}
\affiliation{Department of Physics, University of Virginia, 382 McCormick Rd, Charlottesville, VA 22904-4714, USA}
\author{Aye Win}
\altaffiliation{Now at: Department of Physics, University of Oklahoma, 440 W. Brooks St., Norman, OK 73019, USA}
\affiliation{Department of Physics, University of Virginia, 382 McCormick Rd, Charlottesville, VA 22904-4714, USA}
\author{Miller Eaton}
\affiliation{Department of Physics, University of Virginia, 382 McCormick Rd, Charlottesville, VA 22904-4714, USA}
\author{Niranjan Sridhar}
\altaffiliation{Now at: Google, Inc., 1600 Amphitheatre Parkway, Mountain View, CA, USA}
\affiliation{Department of Physics, University of Virginia, 382 McCormick Rd, Charlottesville, VA 22904-4714, USA}
\author{Reihaneh Shahrokhshahi}
\altaffiliation{Now at: Xanadu, 372 Richmond Street West, Toronto, Ontario M5V1X6, Canada}
\affiliation{Department of Physics, University of Virginia, 382 McCormick Rd, Charlottesville, VA 22904-4714, USA}
\author{Thomas Gerrits}
\affiliation{National Institute of Standards and Technology, 325 Broadway, Boulder, CO 80303, USA}
\author{Adriana Lita}
\affiliation{National Institute of Standards and Technology, 325 Broadway, Boulder, CO 80303, USA}
\author{Sae Woo Nam}
\affiliation{National Institute of Standards and Technology, 325 Broadway, Boulder, CO 80303, USA}
\author{Olivier Pfister}
\affiliation{Department of Physics, University of Virginia, 382 McCormick Rd, Charlottesville, VA 22904-4714, USA}

\begin{abstract}
The Wigner quasiprobability distribution of a narrowband single-photon state  was reconstructed by quantum state tomography using photon-number-resolving measurements with transition-edge sensors (TES) at system efficiency 58(2)\%.  This method makes no assumptions on the nature of the measured state, save for the limitation on photon flux imposed by the TES. Negativity of the Wigner function was observed in the raw data without any inference or correction for decoherence.
\end{abstract}

\maketitle

\section{Introduction}
Single and multiphoton sources prepared in Fock states are of fundamental importance: not only do they enable experiments that epitomize the wave-particle ``duality'' of quantum mechanics, they also can only be described by quantum theory due to the non-positivity of their Wigner quasi-probability distribution.~\cite{Leonhardt1997, Walls1994}

Eugene Wigner originally defined the continuous phase-space quasiprobability distribution function to study quantum corrections to classical statistical  systems \cite{Wigner1932}. For a  quantum state of density operator $\hat{\rho}$, the Wigner function is given by
\begin{align}
W(q,p) &= \frac{1}{\pi}\int_{-\infty}^{\infty}e^{2ipy}\bra{q- y}\hat{\rho}\ket{q+ y}dy,
\end{align}
where $q$ and $p$ are the respective eigenvalues of the position and momentum operators or, in our case, of the amplitude-quadrature, $Q=(\hat{a}+\hat{a}^\dag)/\sqrt2$, and phase-quadrature, $P=i(\hat{a}^\dag-\hat{a})/\sqrt2$, of the quantized electromagnetic field, $\hat{a}$ and $\hat{a}^{\dag}$ being the photon annihilation and creation operators, respectively. The Wigner function is uniquely defined  and contains all the information about the quantum system. It is normalized over phase space and its marginal distributions correspond to the probability density distributions of the quadratures
\begin{align}
    \int_{-\infty}^{\infty}dp\, W(p,q) &= |\psi(q)|^2 \\ \int_{-\infty}^{\infty}dq\, W(p,q) &= |\psi(p)|^2.
\end{align} 
However, unlike classical distributions, the quantum Wigner distribution $W(q,p)$ can't always be interpreted as a joint probability distribution because it can be nonpositive (hence non-Gaussian for pure states~\cite{Hudson}), e.g. for Fock states with $n>0$. This acquires major significance in the context of quantum information and quantum computing over continuous variables (CVQC)~\cite{Lloyd1999,Bartlett2002} as it is well known that  all-Gaussian- (gates and states) CV quantum information suffers from no-go theorems for Bell inequality violation~\cite{Bell1987}, entanglement distillation~\cite{Eisert2002}, and quantum error correction~\cite{Niset2009}. However, none of these no-go theorems apply to CVQC when including non-Gaussian states or gates~\cite{Gottesman2001,Menicucci2006,Menicucci2014ft}.  Non-Gaussian resources are therefore essential to CVQC and can be implemented, for example, by Fock-state generation or detection~\cite{Ghose2007}. 
It is therefore important to be able to characterize Fock states fully and efficiently, possibly in real time. One standard method of state tomography is Wigner function reconstruction. 
Quantum state tomography in phase space~\cite{Smithey1993} can be performed by reconstructing the Wigner function from the measurement statistics of the generalized quadrature ${Q}\cos\phi+ {P}\sin\phi $, measured by balanced homodyne detection (BHD) where phase $\phi$ is the tomographic angle.  This was first done for heralded single photon states in 2001~\cite{Lvovsky2001} and recently improved~\cite{Morin2012}.

\begin{figure}[htb]
\centerline{\includegraphics[width=.5\columnwidth]{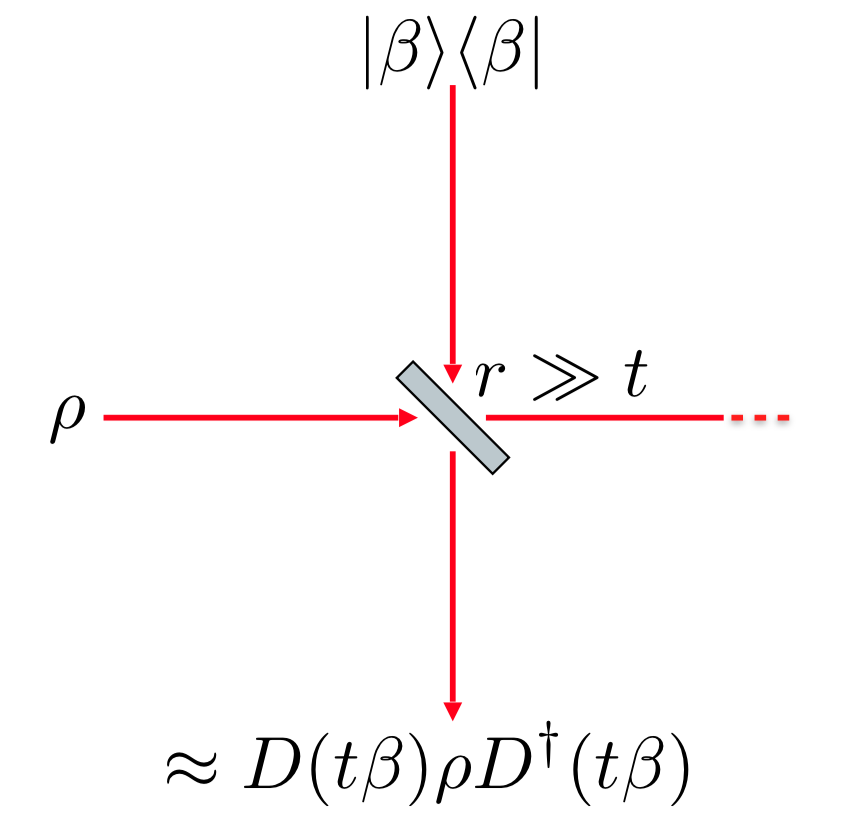}}
\caption{Implementation of a displacement by a beamsplitter. The initial coherent state amplitude is $\beta$ with $|\beta|\gg1$, so that we can have  $t\ll1$ in order to preserve the purity of the quantum signal $\hat{\rho}$, while still retaining a large enough value of $|\alpha|=t|\beta|$, as needed for the raster scan of the Wigner function in phase space.}
\label{fig:disp}
\end{figure}

An issue with BHD-based tomography is that the reconstruction process is computationally intensive, using the inverse Radon transform, or maximum likelihood algorithms~\cite{Lvovsky2009}. 
A more direct approach to reconstruct the Wigner function was proposed by Wallentowitz and Vogel~\cite{Wallentowitz1996} and by Banaszek and Wodkiewicz~\cite{Banaszek1996}. It is based on the following expression of the Wigner function~\cite{Royer1977}
\begin{equation}\label{eq:wigd}
W_{{\hat{\rho}}}(\alpha) = \frac{1}{\pi}Tr[{\hat{\rho}}\hat{D}(\alpha)(-1)^{\hat{N}}\hat{D}^\dagger(\alpha)], 
\end{equation}
where $\alpha =(q+ip)/\sqrt{2}$, $\hat{D}(\alpha)=\exp(\alpha \hat{a}^{\dag}-\alpha^{*} \hat{a})$ is the displacement operator, and $\hat{N}=\hat{a}^{\dag}\hat{a}$ is the number operator. \Eq{wigd} reveals that the Wigner function at a particular phase space point $\alpha$ is the expectation value of the displaced parity operator $\hat{D}(-1)^{\hat{N}}\hat{D}^{\dag}$ over $\hat{\rho}$ or, equivalently, the expectation value of the  parity operator $(-1)^{{N}}$ over the displaced density operator $\hat{D}^{\dag}\hat{\rho} \hat{D}$. This provides a direct measurement method given that one has access to photon-number-resolving (PNR) measurements. In particular, the value of the Wigner function at the origin is the expectation value of the photon number parity operator
\begin{equation}\label{eq:wigd1}
W(0) = \frac{1}{\pi} \sum_{n}^{\infty}(-1)^n\rho_{nn}. 
\end{equation}
Hence, the PNR detection statistics of a quantum system of density operator $\hat{\rho}$  yield a direct determination of the Wigner function at the origin. In order to recover the Wigner function at all points, one can simply displace $\hat{\rho}$ by a raster scan of complex number $\alpha$. This can be done by interference at a highly unbalanced beamsplitter~\cite{Paris1996} of transmission to refelection coefficient ratio $t/r\ll1$, as depicted in \fig{disp}. 
This technique is commonplace in quantum optics and was used, for example, to implement Bob's CV unitary in the first unconditional quantum teleportation experiment~\cite{Furusawa1998}. 
In all rigor, the resulting Wigner function is a more general one, the $s$-ordered Wigner function, $W(s\alpha;s)$, which tends toward $W(\alpha)$ when $s=-t/r\to0$~\cite{Cahill1969}.
This method was implemented for quantum state tomography of phonon Fock states of a vibrating ion~\cite{Leibfried1996}, as well as microwave photon states in cavity QED~\cite{Haroche2002,Vlastakis607}.  For quantum states of light, it has  been experimentally realized for the positive Wigner functions of  vacuum and coherent states, as well as phase-diffused coherent-state mixtures, initially detecting  no more than one photon~\cite{Banaszek1999a} and subsequently detecting several photons~\cite{Bondani2009,Sridhar2014a}.  The nonpositive Wigner function of a single-photon state was reconstructed using PNR measurements by time-multiplexing non-PNR, low efficiency avalanche photodiodes, albeit with the use of {\em a priori} knowledge of the input state in order to deconvolve the effect of losses~\cite{Laiho2010}. 

Our work is the first demonstration of state-independent photon-counting quantum state tomography of a nonpositive Wigner function. The only assumption made here is that the initial quantum state consists of low photon numbers to avoid the saturation limit of the detector, which is less than five photons per microseconds for the superconducting Transition Edge Sensor (TES) used in our experiment.  Since no other prior knowledge is assumed about the state to be measured, this technique is equally applicable to any arbitrary quantum state with low photon flux. We directly observe negativity of the Wigner function with no correction for detector inefficiency. 

\section{Experimental setup and methods} 
 
\subsection{Cavity-enhanced narrowband heralded single-photon source} 
 
\subsubsection{General model}

Our single photon source is based on type-II spontaneous parametric downconversion (SPDC) in a periodically poled $\rm KTiOPO_4$ (PPKTP) crystal. A pump photon at $\om p$  is downconverted into a cross-polarized signal-idler photon pair at $\omega_{s,i}$, such that $\omega_{p}=\omega_{s}+\omega_{i}$, and the presence of the signal photon is heralded by detecting the idler photon~\cite{Lvovsky2001}. All tomographic measurements were therefore conditioned to the detection of an idler photon.
 The SPDC Hamiltonian is given by~\cite{Grice1997,Christ2012}
\begin{equation}\label{eq:PDCH}
H \propto i\hbar\chi^{(2)}\int d^3\vec r\ {E}^{(-)}_p(\vec{r},t) {E}^{(+)}_i(\vec{r},t) {E}^{(+)}_s(\vec{r},t)+ H.c.
\end{equation}
where $\chi^{(2)}$ is the crystal's nonlinearity and the fields in the Heisenberg picture take the form,
\begin{align}
E^{(-)}_{j= p, s,i}(\vec{r},t) &= E^{(+)}(\vec{r},t)^{\dag}\nonumber\\ & = \int d\om j\ A (\vec r,\om j)\, \hat{a}_j\,e^{i[k_j(\om j)r-\om jt]},
\end{align}
where $A (\vec r,\om j)$ is an approximately slowly varying amplitude and $ \hat{a}_j$ is the annihilation operator for the mode of frequency $\om j$. 
Solving for the state under the evolution of the Hamiltonian in \eq{PDCH} for low parametric gain regime and a non-depleted classical pump yields the output quantum state
  \begin{equation}\label{eq:state}
   |\psi\rangle  = \int d^{3}\vec k_{s,i}\,d\om{s,i}\,\phi(\vec{k}_s, \om s,\vec{k_i},\om i) \,\hat{a}^\dagger_s \hat{a}^\dagger_i\ket0_s\ket0_i
  \end{equation}
  where $\phi(\vec{k}_s, \om s,\vec{k_i},\om i)$  determines the spectral and spatial properties of the SPDC, depending on the pump field and the non-linear crystal (phase matching bandwidth around $\vec{k}_p = \vec{k}_s+ \vec{k}_i$). We can see from \eq{state} that the signal and idler photon pairs are emitted in a multitude of spatial and spectral modes.  
Therefore, any measurement on a particular idler mode will collapse the quantum state given by \eq{state} to a mixture of signal-mode states. As a result, the heralded signal state will not be a pure quantum state, which limits its applications in quantum information processing~\cite{Gisin2007, Clauser1978}. This is because a nonzero  vector phase-mismatch can lead to a detected, heralding idler photon with a ``twin'' signal photon completely out of alignment and therefore undetectable, even in the absence of losses, which greatly diminishes the experimentally accessible quantum correlations. One therefore needs to emit photon pairs in the well defined spatial and spectral modes which are optimally coupled to the detectors. This involves spectral and spatial filtering and has been widely studied both theoretically and experimentally~\cite{Ou1999,JeronimoMoreno2010,Virally2010,Scholz2009,Hockel2011,Bao2008}.  Our spectral and spatial filtering steps are discussed in the next section. 

\subsubsection{Spectral and spatial filtering}

Spectral and spatial filtering was achieved by using optical resonators: both actively,  by placing the nonlinear crystal in a resonant cavity --- thereby building an optical parametric oscillator (OPO) --- and passively, by using a filtering cavity (FC) and an interference filter (IF) after the OPO. The OPO was used in the well-below-threshold optical parametric amplifier (OPA) regime. The OPO cavity enhanced the SPDC at doubly resonant (signal and idler) frequencies by a factor of the square of the cavity finesse~\cite{Kuklewicz2002}. However, this enhancement was still masked by the ``sea'' of nonresonant SPDC photons until we filtered the idler with a short FC, which selected only one OPO mode, and with an IF, which selected only one of the FC modes. 
After filtering, we are allowed to consider the simpler OPA Hamiltonian
 \begin{equation}\label{eq:hamil}
H = i\hbar\kappa\, \hat{a}_s^\dagger \hat{a}_i^\dagger+ H.c.,
\end{equation} 
where $\kappa$ is the product of the pump amplitude and $\chi^{(2)}$. This yields the well-known two-mode squeezed state
\begin{equation}\label{eq:TMSV}
|\psi\rangle = (1-\zeta^2)^{\frac12}\sum_{n=0}^{\infty}\zeta^n|n\rangle_s|n\rangle_i,
\end{equation}
where $\zeta=\tanh(\kappa t)$. In the weak pump regime, both $\kappa t$ and $\zeta\ll1$, and \eq{TMSV} can be approximated by
\begin{equation}\label{eq:TMSV1}
|\psi\rangle \simeq 
|0\rangle_s|0\rangle_i + \zeta|1\rangle_s|1\rangle_i.  
+  \mathcal{O}(\zeta^2)
\end{equation}
A detection of a single photon in the idler mode thus projects the signal mode into a single-photon state. 

Note that, since the heralding process consists in postselection of the idler channel, filtering losses in this channel are unimportant. Indeed, if the pump power is kept low enough that practically no pairs from different modes ever overlap in time, one can then reasonably claim that the detected, the heralded signal photon will be the twin of the filtered, heralding idler photon, as per \eq{TMSV1}. 

It is important to also note that the situation will change if one seeks to herald a multi-photon state by using PNR detection for heralding, as per \eq{TMSV}. In that case, losses in the heralding channel cannot be tolerated as they will lead to errors. 

A significant contribution to photon loss is any mode mismatch between the OPO and the FC, which must also be locked on resonance simultaneously, as detailed in the next section. By careful modematching of a seed OPO beam to the FC, we were able to achieve $83\%$ transmission of the OPO mode through the FC. 

\subsection{Experimental setup for quantum tomography }\label{sec:expt}

\subsubsection{Setup description}

\begin{figure*}[htb!]
\centerline{\includegraphics[width=1\textwidth]{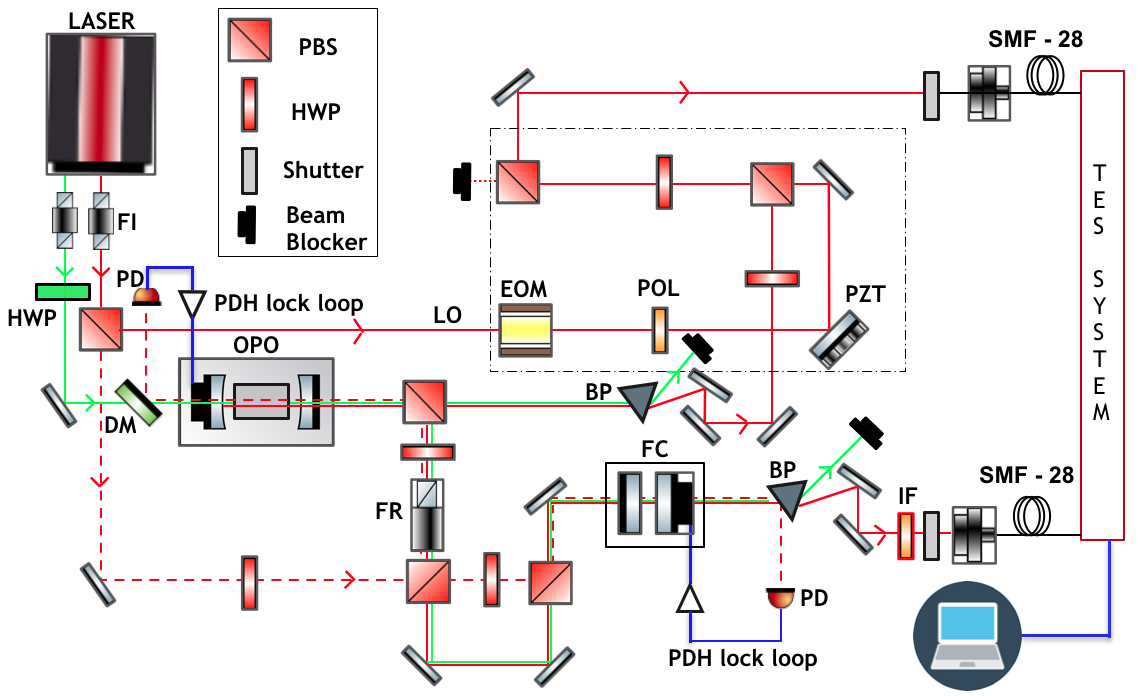}}
\caption{Experimental setup. The red dotted lines denote the locking beam paths for the on/off Pound-Drever-Hall (PDH) servo loops of the OPO and the FC. The displacement operation is contained in the black dash-dotted box at the top.  BP: Brewster prism. DM: Dichroic Mirror. EOM: Electro-optic modulator.  FI : Faraday Isolator. FR: Faraday Rotator. HWP: Half-wave plate. IF : Interference Filter. LO: Local Oscillator input to the displacement field. PBS: Polarizing Beamsplitter. PD: Photodiode. POL: Polarizer. PZT: Piezoelectric transducer.}
\label{fig:newset}
\end{figure*}

The experiment, depicted in \fig{newset}, built upon our previous demonstration of coherent-state tomography~\cite{Sridhar2014a} with the addition of the heralded single-photon source. The OPO was pumped by a stable frequency-doubled 532 nm Nd:YAG nonplanar ring oscillator laser (1 kHz FWHM). A type-II (YZY) quasi-phasematched PPKTP  crystal, of period 450 $\mu$m, was used in the doubly resonant OPO. The two-mirror OPO cavity, as mentioned above, was one-ended, with a finesse of F $\simeq$ 300, 
 an FSR of 1.5 GHz and a FWHM of 5 MHz. One mirror's inside  facet was 99.995\% reflective for the signal and idler fields near 1064 nm and 98\% transmissive for the pump field at 532 nm (the outside mirror facet was uncoated); the other mirror's inside  facet was 98\% reflective at 1064 nm and 99.95\% reflective at 532 nm (its outside facet was antireflection coated at 1064 nm). The cavity was near-concentric with a super-Invar structure, the mirrors' radius of curvature being 5 cm and the mirrors $\simeq$10 cm apart. The FC was made of two 5 cm-curvature, 99\% reflective mirrors placed $\simeq$0.5 mm apart. 
The OPO mode was  aligned and mode-matched to all parts of the experiment (FC, TES fibers) by using a seed beam which was injected into the OPO through its highly (99.995\%) reflecting mirror and exited through its output coupler. The seed beam was carefully mode-matched to the OPO so as to be a pure $\rm TEM_{00}$ mode before being sent to the rest of the setup. It was also used for interference visibility optimization with the displacement field.

\subsubsection{Stabilization procedure}

\begin{figure}[ht!]
    \centerline{
\includegraphics[width=0.5\textwidth]{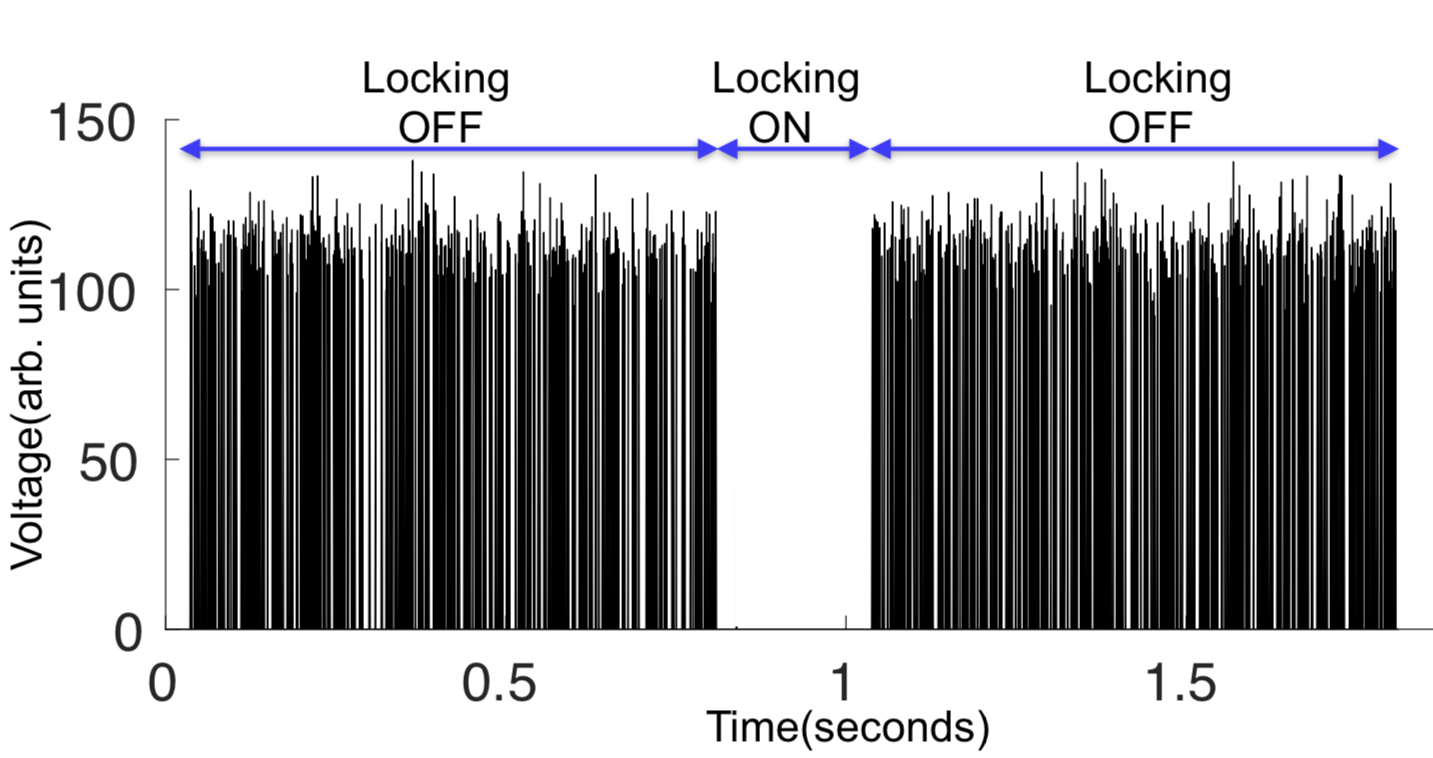}}
    \caption{On/off cycles of alternated active locking and data acquisition. Data collection begins for a period of 800ms while the auxiliary locking beam is blocked, followed by a period of 200ms where the auxiliary locking (broken red line in the experiment schematic) is enabled for active locking and the signal channel is blocked.  This process occurs cyclically during data collection to prevent excess photon flux from damaging the TES while ensuring a stable OPO cavity mode.}
    \label{fig:lock}
\end{figure}

Both the OPO and the FC cavities were Pound-Drever-Hall (PDH) locked~\cite{Drever1983} to a reference laser beam provided by the undoubled output of the pump laser. This was achieved by way of an ``on/off'' locking system, effected by a system of computer-controlled diaphragm shutters. In the ``on'' locking phase, the input to the single-photon sensitive PNR detectors was closed and the reference laser was unblocked and sent into both the OPO and the FC (dotted lines in \fig{newset}) whose PDH lock loops were closed for a few seconds. Because of its super-Invar structure, the OPO drift was low and the PDH loops could then be open, in the ``off'' phase, with their correction signals held constant. The shutter of the reference laser was closed and the paths between the OPO and the PNR detectors were open for data acquisition,  for as long as 3 seconds, see \fig{lock}. 
This procedure allowed us to lock the OPO to its doubly resonant, frequency degenerate mode at $\om s=\om i=\om p/2$. This was essential as the displacement field, also provided by the undoubled output of the pump laser, had to be at the same frequency as the OPO's quantum signal beam and phase-coherent with it. Note that finding this frequency degenerate, doubly resonant OPO mode is nontrivial since the double resonance condition
\begin{align}
\om s &= \om i \\
\Leftrightarrow \frac{m_{s}}{L + n_{s}(T)\ell} &= \frac{m_{i}}{L + n_{i}(T)\ell}
\label{eq:DR}
\end{align}
features two different indices of refraction $n_{s}\neq n_{i}$ ($L$ is the cavity length in air only, $\ell$ the crystal length and $m_{s,i}\in \mathbb{N} $ are the mode numbers). It is, however, possible to temperature-tune the OPO crystal to achieve stable frequency degeneracy~\cite{Feng2003,Feng2004,Feng2004b}. This required temperature control of the PPKTP crystal to the level a few millidegrees, around $27.810^{\circ}$ C, using a commercial temperature controller.
 
\subsubsection{PNR detection}

Our PNR detection system is comprised of two transistion edge sensors (TES), consisting of tungsten chips in a cryostat, coupled through standard telecom fiber. A detailed description of the TES system can be found in Refs.~\citenum{Lita2008,Sridhar2014a}. The TESs are cooled using an adiabatic demagnetization fridge at a stable temperature of 100 mK, at  the bottom edge of the steep superconducting transition slope (resistance versus temperature). When one photon is detected, its energy is absorbed by the tungsten chip,  yielding a sharp increase in its  resistance which is detected by a SQUID over a rise time on the order of 100 ns. The heat is then dissipated through a weak thermal link, over a time on the order of 1 $\mu$s. During this time, the TES is still active (as opposed to, say, of nanowire detectors or avalanche photodiodes).  Due to the finiteness of its superconducting transition slope, the TES can resolve up to 5 photons. The absolute maximum photon flux sustainable by the TES without the tungsten driven into the normal conductive regime is therefore 5 photons/$\mu$s in the continuous-wave regime, i.e., a power of 1 pW. The OPO's average power was kept at 100 fW by setting the pump power to 200 $\mu$W (the OPO threshold was 200 mW). We observed that the background counts were negligible when the TES signal was suppressed by rotating the pump's linear polarization by 90$^{\circ}$, thereby completely phase-mismatching the nonlinear interaction in PPKTP.
\subsubsection{Displacement calibration}\label{sec:calibration}

The displacement operator was implemented by interfering the OPO signal mode with a phase- and amplitude-shifted coherent-state displacement field at a highly unbalanced beamsplitter with a reflectivity  $r^2=0.97$. The interference visibility between the seed OPO beam and the displacement field was 90\%. The  amplitude shift $|\alpha|$ was effected by a homemade, temperature-stabilized electro-optic modulator consisting in an X-cut, 20 mm-long rubidium titanyl arsenate ($\rm RbTiOAsO_{4}$) crystal; the phase shift $\arg(\alpha)$  was effected by a piezoelectric transducer- (PZT) actuated mirror. Both the EOM and the PZT mirror were driven by homemade, low-noise, high voltage drivers, fed by computer-controlled lock-in amplifiers. 

The amplitude displacement was varied in 20 steps from $|\alpha|=0$ to $|\alpha_\text{max}| =  0.796(7)$, fixed by the TES' photon flux limit of 5 photons/$\mu$s. The phase displacement was varied in 10 steps from 0 to $2\pi$. The amplitude steps $|\alpha|=\sqrt{\eta}|\beta|$, where $\eta$ is the overall detection efficiency, were directly calibrated by comparing the TES photon statistics to that of a Poisson distribution 
\begin{equation}
P(n) = e^{-|\alpha|^2}\frac{|\alpha|^{2n}}{n!},
\end{equation}
with the OPO beam blocked. This allowed us to determine the displacement amplitude 
\begin{equation}
|\alpha| = \left[\frac{2P(2)}{P(1)}\right]^{\frac12}.
\end{equation}
Note that this method requires the presence of 2-photon detection events, i.e.,  $|\alpha_{\text{min}}| \simeq 0.15$ for the very first displacement amplitude, besides the zero displacement for which we blocked the displacement beam. Photon number statistics were averaged over 2 seconds to ensure an average calibration accuracy  
\begin{equation}
\triangle |\alpha| = 3\times10^{-3}
\end{equation}
of the displacement amplitude. However, the error on the maximum displacement was somewhat larger 
\begin{equation}
\triangle |\alpha_\text{max}| = 7\times10^{-3},
\end{equation}
due to the photon pileups occurring at higher flux which make the continuous-wave TES signals harder to analyze. 
We observed the long-term power stability of the laser to be on the order of 1\% over an hour.  The laser's short-term intensity noise was much lower as ensured by a built-in ``noise eater'' intensity servo. Moreover, the temperature stability of the EOM was on the order of 1 mK.  Because of all this, we consider the error $\triangle |\alpha|$ on the displacement calibration to be valid over the course of our data acquisition time of several minutes. 

The phase steps were calibrated by scanning the interference fringe between the OPO seed beam and the displacement field, which provided a set of 10 voltage values for the PZT mirror. Experimental data runs were conducted by scanning the amplitude at fixed phases, with the phase PZT voltage being refreshed at every amplitude EOM voltage step.  For each point of the quantum phase space, a continuous stream of data was acquired at 5 MS/s, digitized using an PCI board, and stored for subsequent photon statistical analysis. A detailed discussion of our data analysis of  continuous-wave photon counting can be found in our previous paper on coherent state tomography using PNR measurements~\cite{Sridhar2014a}. 

\section{Experimental results}\label{sec:results}

\subsection{Heralding ratio}

The heralding ratio determines the quality of the single-photon source. It is the probability of seeing one photon in the OPO  signal (heralded) beam with no displacement field, provided one photon was detected in the filtered  idler (heralding) beam. The pump power was kept low enough so as to suppress two-photon events in the OPO signal in the absence of a displacement field. Results are displayed in \tab{label}.  
\begin{table}[h!]
\[
\begin{array}{c|c|c|c}
  \hline
  & N_s & N_i & N_c \\
  \hline 
\text{Single-photon events} & 54320\pm90 & 1556\pm30 &  903\pm17\\
  \hline
\end{array}
\]
\caption{Experimentally measured number of single-photon counts on both channels. $N_s$: number of photons in the (heralded) signal channel, $N_i$: number of  photons on the (heralding) idler channel,  $N_c$: number of coincident counts. }
\label{tab:label}
\end{table}
We can see that the heralded channel has a lot more counts than the heralding channel, as expected since the latter is filtered by the FC and the IF. The heralding efficiency  was
\begin{align}\label{eq:k} 
\eta_{h} &= \frac{N_c}{N_i}  \\
& = 0.58\pm0.02 
\end{align}
and can also be considered the overall detection efficiency of the heralded channel, i.e., of the quantum signal. 

\subsection{Photon probability distributions versus displacement amplitude}
\begin{figure}[h!]
  \centerline{\includegraphics[width=0.5\textwidth]{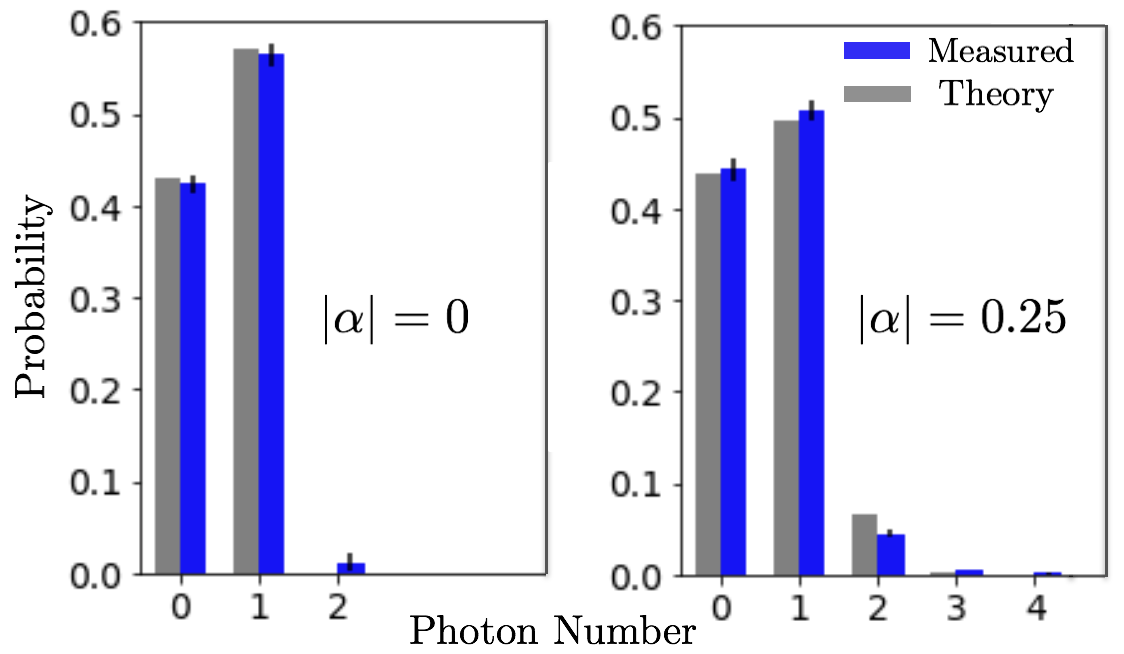}}
    \caption{Measured photon-number distributions, left: $\alpha = 0$ and right : $|\alpha| = 0.25$. Error bars (1$\sigma$) are calculated from the statistics of the measurements. }
    \label{fig:hist}
\end{figure}

\Fig{hist} displays the measured photon number distributions for a heralded single-photon input when  $|\alpha|$ = 0 (left) and 0.25 (right).  For no displacement, the histogram reflects the exact same measurement as in \tab{label} and \eq k, and the result yields a compatible value of $0.58(2)$. The two-photon counts are essentially absent, which results in a very low second-order coherence $g^2(0) =  0.07(5)$. For $|\alpha|$ = 0.25, the two-photon peak grows from the presence of the displacement field. In both cases, the observations agree with the theoretical distribution, calculated with $\eta$= 0.58. 
As expected, the single-photon component decreased while the vacuum and higher photon components increased. It can be thought of as follows: If we displace a pure single-photon state, then we obtain
\begin{align*}
 D(\alpha)|1\rangle &=D(\alpha)a^{\dag}D^{\dag}(\alpha)|\alpha\rangle\\
&=a^{\dag}|\alpha\rangle-\alpha^{*}|\alpha\rangle. 
\end{align*}
Clearly, the first term has no vacuum component, as does the initial state $|1\rangle$, but the second term does have a vacuum component. Therefore, the displacement of a single-photon state increases its vacuum component probability amplitude, somewhat unintuitively.  In the case where our initial state is a mixture of vacuum and single-photon, then it can be seen that for low enough displacement amplitudes, the vacuum component still increases from its previous value.  As the displacement becomes large, the vacuum component will eventually decrease.

\subsection{Model Wigner function and loss analysis}
Before we turn to the tomography results, we outline the Wigner function model that accounts for the aforementioned nonideal system detection efficiency. 
There are several sources of losses in our experiment: photon absorption and general scattering out of the mode due to mismatch. As mentioned above, losses in the heralding channel can be factored out in the generation of a heralded single-photon state provided that the OPO output never contains more than one photon per mode during the detection window, which was the case in this work. 

 It is also important to note that the TES fiber is single-mode at telecom wavelengths but not at our operating wavelength of 1064 nm. Hence we need to address the possibility of  multimode coupling into the TES fiber. A simple reasoning shows that this is not a matter of concern if there are no losses in the fiber. Indeed, the coupling of the input field into each of the different, orthogonal propagation modes of the fiber can be accurately described by as many beamsplitting operations into distinguishable outputs. While each of these beamsplitting operations does bring in vacuum fluctuations, all beamsplitter outputs are still detected and the final TES detection is simply that of the total photon number of all the fiber modes. In the absence of losses, the multimode fiber is a passive optical element which conserves the total photon number and the final total photon number measurement must therefore give the same exact result as the initial one, before the quantum light is coupled into the fiber. 
 An argument could be made that fiber losses could be mode dependent, with higher-order modes being more likely to leak out of the fiber; we assume that this is negligible in our case because the operating wavelength was close enough to the specified single-mode wavelength that the mode order should not be that high. 

\begin{figure}[ht]
  \centerline{\includegraphics[width=0.4\textwidth]{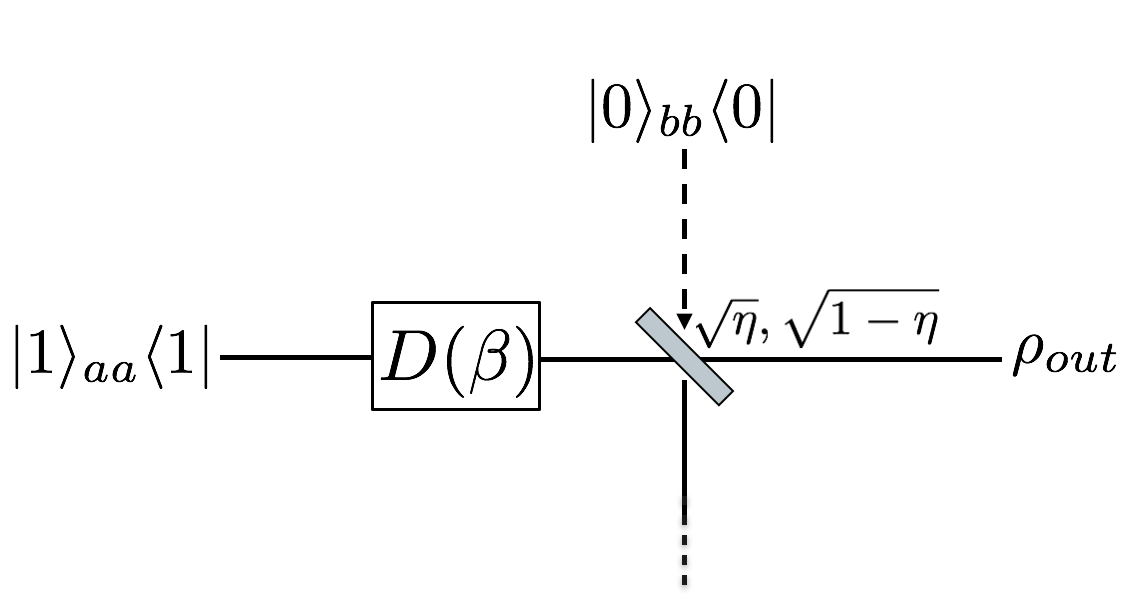}}
    \caption{Loss model. The beamsplitter transmission and reflection coefficients are $\sqrt{\eta}$ and $\sqrt{1-\eta}$ respectively. }
    \label{fig:loss}
\end{figure}

We measured the coupling efficiency, $\eta_\text{OFC}$, into the TES fiber on the optical table by cleaving the fiber to insert a power meter and re-fusing it to the TES thereafter. To minimize the coupling to higher modes, we optimized our fiber coupling to as high as 90\% with the seed beam (discussed in the ``spectral and spatial filtering'' section above) and we also measured the intensity variations of about one percent at the output of the fiber. This ensures that most of the fiber coupling was to the fundamental mode of the fiber. However, we didn't measure the overall fiber transmission into the TES cryostat. This was bundled with the TES quantum efficiency in $\eta_\text{TES}$, which was inferred from all other measured efficiencies, as summarized in \tab2. 
\begin{table}[h!]
\[\begin{array}{cccc|c}
\hline
\eta_\text{TES}  & \eta_\text{OT} & \eta_\text{BS} &\eta_\text{OFC} & \eta \\
\hline
0.71(3) & 0.93(1)& 0.97(1)& 0.90(2) &0.58(2)\\
\hline
\end{array}\]
\caption{$\eta_\text{TES}$: TES quantum efficiency (including fiber transmissivity); $\eta_\text{OT}$ : optical transmission of single-photon signal field from the OPO to the displacement operation; $\eta_\text{OFC}$: optical fiber coupling. The overall efficiency $\eta= \eta_\text{TES}  \times \eta_\text{OT} \times \eta_\text{BS} \times\eta_\text{OFC}$.}
  \label{tab:2}
\end{table}
We modeled  losses by considering a fictitious beamsplitter of transmissivity $\eta$ and reflectivity $(1-\eta)$, placed between the displacement and a detector of unity efficiency as shown in \fig{loss}. The input state of this system is
\begin{equation}
\hat{\rho}_\text{in} = |1\rangle_a\,{}_a\!\langle1|\otimes|0\rangle_b\,_b\!\langle0|.
\end{equation} 
After applying the displacement $\hat{D}_{a}(\beta)$ and beamsplitter $\hat{U}_{ab}$ operators we obtain the reduced, detected density operator by tracing out the vacuum mode
\begin{align}\label{eq:rhobs}
\hat{\rho}_\text{out} &= \tr b{\hat{U}_{ab} \hat{D}(\beta)|1\rangle_a\,{}_a\!\langle1|\otimes|0\rangle_b\,_b\!\langle0| \hat{D}^\dagger(\beta) \hat{U}_{ab}^\dagger}\\
&=  \hat{D}(\sqrt\eta \beta) \left[\eta|1\rangle_a\,{}_a\!\langle 1|  + (1-\eta) |0\rangle_a\,_a\! \langle 0|\right]  \hat{D}(\sqrt\eta \beta)^\dagger \label{eq:rhor}.
\end{align} 
From \eq{rhor} we can see that displacement by $\beta$ followed by losses $\eta$ is essentially the same as introducing losses first by mixing the pure single-photon state with vacuum, and then applying a displacement by the reduced amount $\sqrt{\eta}\beta$. Due to the linearity of the Wigner function, \eq{rhor} shows that the experimentally reconstructed Wigner function will in fact be
\begin{equation}\label{eq:fit}
W(p,q) = \eta W_{|1\rangle \langle1|}(p,q) + (1-\eta )W_{|0\rangle \langle0|}(p,q).
\end{equation}
As expected, losses (1-$\eta$) add a Gaussian vacuum function to the original nonpositive Wigner function of the single-photon state. In particular, the undisplaced photon-number distribution will yield the overall transmissivity of the whole experiment $\eta$, as in \fig{hist}, left.
\subsection{Quantum tomography of a single-photon state}
\begin{figure}[htb]
\centerline{\includegraphics[width=1\columnwidth]{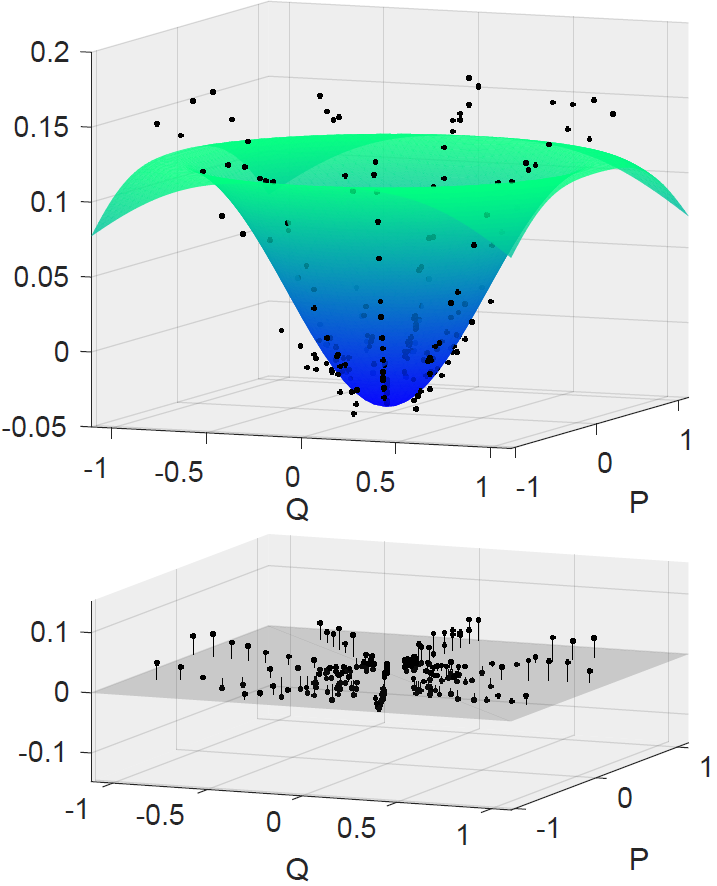}}
\caption{Top, reconstructed Wigner function. Black points: reconstructed values from raw data. Solid surface: least-square Wigner-function fit, \eq{fit}. Bottom, fit residuals.}
\label{fig:surf}
\end{figure}

We now turn to the state tomography results. \Fig{surf} shows the reconstructed Wigner function along with a fit with Wigner function \eq{fit}, of free parameter $\eta$. 
\begin{figure}[hbt]
\centerline{\includegraphics[width=0.45\textwidth]{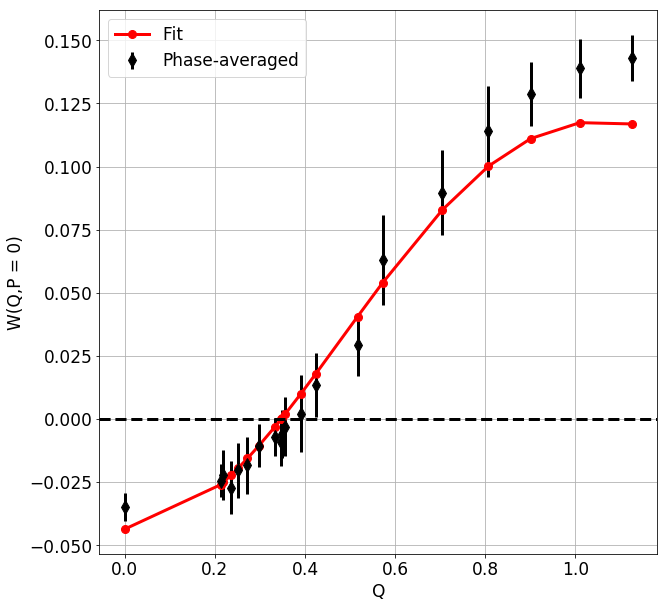}}
\caption{Phase-averaged Wigner function. The Wigner function fit yielded $\eta=0.57(3)$, which is consistent with the heralding efficiency $0.58(2)$. Error bars are discussed in the text.}\label{fig:phase_ave}
\end{figure}
The Wigner function is plotted for experimentally measured values of $|\alpha|$ where phase space coordinates are $(q,p)= (\sqrt{2}|\alpha|\cos\phi,\sqrt{2}|\alpha|\sin\phi)$, where $\phi$ is the tomographic angle. We can clearly see the negativity around the origin of the phase space,
\begin{equation}
 W(0,0) = -0.035\pm 0.005.
\end{equation}
Errors in the displacement amplitudes were considered to be negligible due to the long-term amplitude stability of the laser producing the displacement field and the high-accuracy of the calibration as mentioned in 
the ``displacement calibration'' section. The Wigner function error  bars (1$\sigma$) at zero-displacement were obtained from the statistics of multiple data sets with the displacement field blocked. At non-zero displacement, in order to speed up the measurement process and minimize experimental drifts, we decided to use the statistics of the measurement results at 10 different phases for the same displacement amplitude.  This procedure yields a conservative estimate of the Wigner function error bars (1$\sigma$), in the particular case of a single-photon Fock state, because it assumes that the measured Wigner function has the required cylindrical symmetry about the origin of phase space.  The results are plotted on \fig{phase_ave}.  Note that the fact that Wigner function isn't significantly altered by this averaging --- in fact, both the 2D fit in \fig{surf} and the 1D fit in \fig{phase_ave} yield $\eta=0.57(3)$ --- speaks to the high quality of the phase-space rotational symmetry of our data. 
One can notice that the fit residuals are reasonably small around the origin of phase space but grow larger in the outskirts of the function, near our maximum displacement values. These correspond to larger detected photon numbers on the TES, for which photon pileups during the TES' cooling time make data analysis more arduous~\cite{Sridhar2014a}. 
\section{Conclusion}
We have demonstrated state-independent photon-counting quantum state tomography with PNR measurements using a superconducting TES system and evidenced clear negativity in the single-photon Fock Wigner function with no correction for photon loss. This work has been limited by two factors: when working with continuous-wave detection, photon fluxes become overwhelming to the TES when $|\alpha|\to1$. Moreover, photon pileups, in particular during the TES cooling time, greatly complicate data analysis~\cite{Sridhar2014a}. In the future, we will multiplex several TES channels in order to to access larger displacement amplitudes, i.e., larger regions of phase space. This will also reduce the photon pileup effect. Finally, owing to the intrinsic simplicity of photon-counting quantum tomography, we believe it is possible to herald and visualize Fock state Wigner functions in real time for quantum information applications. 
\section*{Funding Information}
This work was supported by NSF grants PHY-1521083 and PHY-1708023.
\section*{Acknowledgments}
The authors thank  Rafael Alexander, Carlos Andreas Gonzalez Arciniegas, Xu Yi, Avi Pe'er, Chun-Hung Chang, Jacob Higgins, Chaitali Joshi, and Xuan Zhu for helpful discussions.  We would also like to thank Scott Glancy and Arik Avagyan for valuable comments and feedback during the revision process. 

\bibliography{Pfister}
\bibliographystyle{bibstyleNCM}

\end{document}